\documentclass{kluwer}    % Specifies the document style.

\usepackage{psfig}

\def\etal{{et\,al. }}
\newbox\grsign \setbox\grsign=\hbox{$>$}
\newdimen\grdimen \grdimen=\ht\grsign
\newbox\laxbox \newbox\gaxbox
\setbox\gaxbox=\hbox{\raise.5ex\hbox{$>$}\llap
     {\lower.5ex\hbox{$\sim$}}}\ht1=\grdimen\dp1=0pt
\setbox\laxbox=\hbox{\raise.5ex\hbox{$<$}\llap
     {\lower.5ex\hbox{$\sim$}}}\ht2=\grdimen\dp2=0pt

\def\lax{$\mathrel{\copy\laxbox}$}

\begin{document}                                                                                   
\begin{article}
\begin{opening}         
\title{K band monitoring of GRS 1915+105 during 1999/2000
%\thanks{Footnote 
%            to the title with the `thanks' command.}
} 
\author{J. \surname{Greiner}}  
\runningauthor{Greiner et al.}
\runningtitle{IR monitoring of GRS 1915+105}
\institute{Astrophysical Institute Potsdam, 14482 Potsdam, Germany}
\author{F.J. \surname{Vrba}}
\author{A.A. \surname{Henden}}
\author{H.H. \surname{Guetter}}
\author{C.B. \surname{Luginbuhl}}
\institute{US Naval Observatory, Box 1149, Flagstaff Station,
   Flagstaff, AZ 86002, USA}

%\date{April 15, 1993}

\begin{abstract}
We report on long-term K band photometry of the galactic microquasar 
GRS 1915+105 since March 1999. We see variations of more than 1 mag,
which are seemingly correlated with variations in X-ray intensity and spectral 
slope. 
%We discuss a possible explanation of this correlated variability.
\end{abstract}

\end{opening}           

\section{Introduction}  

GRS 1915+105 is the prototypical microquasar, a galactic X-ray binary
ejecting plasma clouds at v$\approx$0.92\,c (Mirabel \& Rodriguez 1994).
%It exhibits unique X-ray variability patterns (Greiner \etal\ 1996) 
%which have been interpreted as
%instabilities leading to an infall of parts of the
%inner accretion disk (Belloni \etal\ 1997).
%  -- a galactic source at distance d $\approx$ 9--12 kpc 
%      (Fender \etal\ 1999)\\
%  -- at bII = -0\fdg2 behind a column of A$_V$ $\approx$ 27 mag \\
%Some correlated X-ray/infrared/radio variability patterns have also been 
%related to jet formation  (Mirabel \etal\ 1998).
Three different sources of infrared (IR) radiation are expected from a 
X-ray binary,
namely thermal emission from the outer part of accretion disk, thermal 
emission from the companion star or synchrotron emission from the jet(s).
%Infrared variability could be caused by a variety of processes, among them
%changes of the mass transfer rate from the secondary to the accretion disk,
%a changing aspect of the illuminated secondary,
%a changing illumination of the disk and/or secondary, 
%sporadic jet ejection and infrared synchrotron emission, or
%free-free emission from an X-ray driven wind.
Previous IR observations of GRS 1915+105 include, besides
a few single measurements since 1993, 
some few hrs photometry in conjunction with X-ray/radio measurements 
%     showing 3 different types of IR-behaviour 
(Eikenberry \etal\ 2000), and
a 2 month monitoring in 1996 (Bandyopadhyay \etal\ 1998).

\section{Observations and Results}

K band
observations have been performed with the 1.55-m telescope of the US Naval 
Observatory Flagstaff
equipped with the 256$\times$256 pixel HgCdTe IRCAM camera
%(USNOFS) 
during 8--10 day intervals around full Moon.
% During these times the telescope is 
%equipped with the
%256$\times$256 pixel HgCdTe IRCAM camera.
% giving a camera scale of 0\farcs54/pixel.
Because the main interest was in the long-term behaviour, ``only''
about 1--3 observations per month were done, consisting of
groups of 3 or 4 individual observations per night. 
Each group represents a set of coadded integrations with an exposure 
of 400 sec. In total, observations were performed on 34 nights between 
05 May 1999 and 15 Jun 2000.

 \begin{figure}[th]
   \centering{
   \hspace{-0.01cm}
   \vbox{\psfig{figure=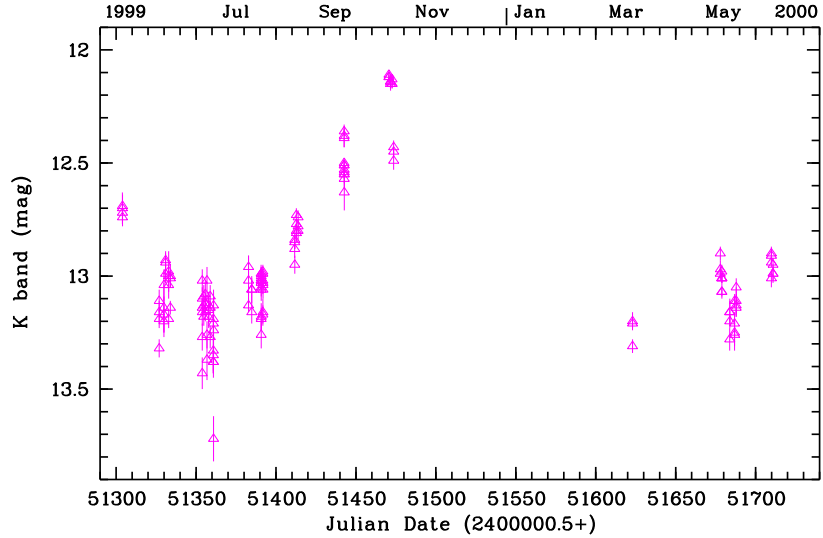,width=0.93\textwidth,%
         bbllx=1.9cm,bblly=1.6cm,bburx=18.2cm,bbury=12.6cm,clip=}}\par
   \vspace{-0.25cm}
   \caption[lc]{K band light curve of GRS 1915+105, showing the mean of
         all ``groups''.
          }}
         \label{lc}
 \end{figure}

%These findings do not contradict the results of 
%\citeauthor{Chin88-book} (on which \shortcite{Bunt} based their 
%production kinetics) and of \citeyear{ChinThesis} which were obtained for

%\section{Results and Discussion}

These 
%long-term monitoring 
observations show GRS 1915+105 to be variable 
%in the K band 
between 12.1 \lax\ K \lax\ 13.7 mag (not extinction-corrected; Fig. 1).
%While this amplitude has been noticed already earlier, 
Our long coverage
allows to search for correlations with other emission characteristics, e.g.
%In particular, we used 
the RXTE All-Sky Monitor (ASM) 
quick-look results provided by the ASM/RXTE team.
After selecting those K band observations which were
done within 4 hrs of an ASM scan, we find that both
the X-ray flux 
%in the 2--12 keV range 
as well as the X-ray hardness ratio (flux ratio of 5--12 keV vs. 3--5 keV band)
are positively correlated with the K band flux. 
%The hardness ratio was computed as the ratio of flux in the
%5--12 keV to 3--5 keV band, and thus is primarily a measure of the slope of 
%the power law component.

A regression analysis 
%between the X-ray K band flux 
suggests that a
doubling of the X-ray count rate (flux) leads to a 33\% increase in the
K band flux.
%(see Greiner \etal\ 2001 for more details). 
%Exceptions to this regression are the two extreme K band values 
%K=13.7 mag and K=12.1 mag which both are simultaneous to strong 
%radio (and X-ray) flaring episodes. 
%The fact that the K band flux does not
%increase at the same rate as the X-ray flux clearly argues against
This argues against
an explanation due to illumination of the secondary. Instead, it rather
could be synchrotron radiation 
%due to the accretion vs. jet-ejection relation 
related to (small) ejections/flares as seen also at
shorter time scales 
%(Mirabel \etal\ 1998, 
(Eikenberry \etal\ 2000).
In this picture
the correlation of the K band flux with the X-ray hardness ratio is 
compatible with the correlation of X-ray spectral slope and radio flux 
found by Rau \& Greiner (2001). A large (small) value of the hardness ratio
corresponds to a flat (steep) X-ray power law slope, and therefore a high (low)
temperature of the electrons in the corona if the general comptonization 
paradigm is adopted. These correlations then imply that strong K band flux
and weak radio flux occur during times of a flat X-ray power law, i.e.
most of the synchrotron power is radiated either in the infrared or radio
band depending only on the temperature of electrons in the corona. 
%Thus, 
%it seems reasonable to assume that the varying infrared and radio emission
%as well as the slope of the X-ray power law are determined by just one
%parameter, namely the electron temperature.

%\acknowledgements
%And this is an acknowledgement

\end{article}
\end{document}